\documentclass[aps,prb,floatfix,reprint,twocolumn,reprint,amsmath,amssymb,superscriptaddress,showpacs]{revtex4-1}
\usepackage{graphicx}
\usepackage{bm}
\usepackage{amsmath}
\usepackage{dcolumn}
\usepackage{dsfont}
\usepackage{amssymb}
\usepackage{tabularx}
\usepackage{array}
\usepackage{float}
\usepackage{color}
\usepackage{epstopdf}
\usepackage{mathrsfs}
\usepackage[colorlinks, linkcolor=blue,anchorcolor=blue,citecolor=blue,urlcolor=blue]{hyperref}

\begin{document}

\title{Two-dimensional Spin-Orbit Dirac Point in Monolayer HfGeTe}

\author{Shan Guan}
\affiliation{Beijing Key Laboratory of Nanophotonics and Ultrafine Optoelectronic Systems, School of Physics,
Beijing Institute of Technology, Beijing 100081, China}
\affiliation{Research Laboratory for Quantum Materials, Singapore University of Technology
and Design, Singapore 487372, Singapore}

\author{Ying Liu}
\affiliation{Research Laboratory for Quantum Materials, Singapore University of Technology
and Design, Singapore 487372, Singapore}

\author{Zhi-Ming Yu}
\affiliation{Research Laboratory for Quantum Materials, Singapore University of Technology
and Design, Singapore 487372, Singapore}

\author{Shan-Shan Wang}
\affiliation{Research Laboratory for Quantum Materials, Singapore University of Technology
and Design, Singapore 487372, Singapore}

\author{Yugui Yao}
\email{ygyao@bit.edu.cn}
\affiliation{Beijing Key Laboratory of Nanophotonics and Ultrafine Optoelectronic Systems,
School of Physics, Beijing Institute of Technology, Beijing 100081, China}

\author{Shengyuan A. Yang}
\email{shengyuan\_yang@sutd.edu.sg}
\affiliation{Research Laboratory for Quantum Materials, Singapore University of Technology
and Design, Singapore 487372, Singapore}


\begin{abstract}
Dirac points in two-dimensional (2D) materials have been a fascinating subject of research, with graphene as the most prominent example. However, the Dirac points in existing 2D materials, including graphene, are vulnerable against spin-orbit coupling (SOC). Here, based on first-principles calculations and theoretical analysis, we propose a new family of stable 2D materials, the HfGeTe-family monolayers, which represent the first example to host so-called spin-orbit Dirac points (SDPs) close to the Fermi level. These Dirac points are special in that they are formed only under significant SOC, hence they are intrinsically robust against SOC. We show that the existence of a pair of SDPs are dictated by the nonsymmorphic space group symmetry of the system, which are very robust under various types of lattice strains. The energy, the dispersion, and the valley occupation around the Dirac points can be effectively tuned by strain. We construct a low-energy effective model to characterize the Dirac fermions around the SDPs.  Furthermore, we find that the material is simultaneously a 2D $\mathbb{Z}_2$ topological metal, which possesses nontrivial $\mathbb{Z}_2$ invariant in the bulk and spin-helical edge states on the boundary. From the calculated exfoliation energies and mechanical properties, we show that these materials can be readily obtained in experiment from the existing bulk materials. Our result reveals HfGeTe-family monolayers as a promising platform for exploring spin-orbit Dirac fermions and novel topological phases in two-dimensions.
\end{abstract}


\maketitle
\section{Introduction}
Graphene, as the most prominent example of two-dimensional (2D) materials~\cite{Novoselov2004}, has been attracting tremendous interest in the past decade. Many of the excellent properties of graphene can be attributed to its Dirac-cone-type band structure~\cite{CastroNeto2009}. The conduction and valence bands in graphene touch with linear dispersion at discrete Dirac points on the Fermi level, around which the low-energy electrons behave like relativistic massless Dirac fermions in 2D, exhibiting properties distinct from the usual Schr\"{o}dinger fermions. Inspired by graphene, much effort has been devoted to the search for other 2D materials which host Dirac/Weyl points, and a number of candidates have been proposed~\cite{Wang2015}, such as silicene~\cite{Cahangirov2009,Liu2011}, germanene~\cite{Cahangirov2009,Liu2015}, graphyne~\cite{Malko2012}, 2D carbon and boron allotropes~\cite{Xu2014,Zhou2014,Ma2016,Jiao2016,Feng2017}, group-VA phosphorene structures~\cite{Kim2015,Lu2016}, and 5$d$ transition metal trichloride~\cite{Sheng2017}. The Dirac points in all these materials (including graphene) are protected by symmetry, but \emph{only} in the absence of spin-orbit coupling (SOC). When SOC is included, a gap will be opened at the Dirac point, so strictly speaking, graphene is formally a 2D topological insulator (also known as quantum spin Hall insulator)~\cite{Kane2005}, although the gap size is very small~\cite{Min2006,Yao2007}.

Is it possible to have \emph{2D Dirac points that are robust against SOC}? This question has been theoretically addressed by Young and Kane~\cite{Young2015a}. Via symmetry analysis and tight-binding model studies, they showed that certain nonsymmorphic space group symmetries, i.e., symmetries involving fractional lattice translations, can stabilize a new kind of 2D Dirac points that are robust against SOC. Besides the stability under SOC as their defining signature, such Dirac points, termed as 2D spin-orbit Dirac points (SDPs), also exhibit the following features that are different from the previously studied (SOC-vulnerable) Dirac points: (i) their presence is solely dictated by the specific space group symmetry and does not require band inversion; (ii) they must locate at time-reversal-invariant momenta at the Brillouin zone (BZ) boundary; and (iii) their appearance close to Fermi level typically requires partial filling of multiple bands.

Despite the recent exciting advance in theory~\cite{Yang2014,Wieder2016} and in finding analogous 3D SDPs in several bulk materials~\cite{Young2012,Steinberg2014}, the search for realistic 2D materials that possess 2D SDPs at low energy is still challenging. The crucial issue is regarding the structural stability. Stable 2D materials with nonsymmorphic space group symmetries are typically insulators. For example, group-VA 2D materials with phosphorene structure~\cite{Lu2016} or 2D materials with $\alpha$-SnO structure~\cite{Singh2014} do possess SDPs in their band structures, but these SDPs are away from the Fermi level (usually by energy $>0.5$ eV), so they will hardly manifest in electronic properties. On the other hand, if one tries to expose these SDPs to Fermi level by replacing the elements with other species of different valence, the resulting 2D materials are usually found to be structurally unstable. So far, a stable 2D material hosting SDPs close to its Fermi level has not been found yet, and how to realize such a material remains an open problem.

In this work, we reveal a family of monolayer materials as the first example of 2D SDP materials, which realize the 2D SDPs proposed by Young and Kane. Using first-principles calculations and theoretical analysis, we show that monolayer HfGeTe-family materials host a pair of 2D SDPs close to the Fermi level. We demonstrate the stability of these materials in the monolayer form, and more importantly, for several of the member materials (including HfGeTe), the corresponding three-dimensional bulk materials already exist and the calculated exfoliation energy is relatively low, so that the monolayer can be more readily obtained, e.g., by mechanical exfoliation from the bulk. From symmetry analysis, we show that the two SDPs are dictated by the nonsymmorphic space group symmetry to appear at high symmetry points on the BZ boundary. An effective model is constructed to characterize the low-energy Dirac fermions. We further show that the SDPs are not only robust against SOC, they also survive under a variety of lattice strains: biaxial, uniaxial, and shear strains all preserve the SDPs; and they serve as effective means to tune the Dirac dispersion as well as the energy of the SDPs relative to the Fermi level. In addition, we find that these materials carry a well-defined $\mathbb{Z}_2$ invariant despite a vanishing global bandgap, corresponding to a 2D $\mathbb{Z}_2$ topological metal. Consequently, they possess topological edge states at the system boundary, which is indeed confirmed by our calculation. Our finding opens the door to the exploration of 2D spin-orbit Dirac materials, and provides a realistic material platform for the fundamental research as well as promising nanoscale applications.

\section{COMPUTATIONAL DETAILS}\label{section:methods}
Our first-principles calculations are based on the density functional theory (DFT), using the projector augmented wave method~\cite{Bloechl1994} as implemented in the Vienna \emph{ab-initio} Simulation Package~\cite{Kresse1993,Kresse1996}. The generalized gradient approximation (GGA) with Perdew-Burke-Ernzerhof (PBE) realization~\cite{PBE} is used for the exchange-correlation functional. Our main results are also verified by the hybrid functional approach (HSE06)~\cite{Heyd2003}. The plane-wave energy cutoff is set to be 330 eV. Monkhorst-Pack $k$-point mesh with size of 30$\times$30$\times$1 is applied for the Brillouin zone sampling. A vacuum layer of 13 \AA\, thickness is added to avoid artificial interactions between periodic images. All lattice structures are fully relaxed until energy and force are converged with accuracy of $10^{-6}$ eV and 0.005 eV/\AA, respectively. Van der Waals interaction is taken into account by using the approach of Dion \emph{et al}~\cite{Dion2004}.  The phonon spectrum is calculated using the PHONOPY code through the DFPT approach~\cite{Togo2015}, with a $6\times 6\times 1$ supercell and a $4\times 4\times 1$ $q$-grid (such that the total energy converges with an accuracy of $10^{-8}$ eV). The transition metal $d$ orbitals may have
important correlation effects, so we also validate our results by using the DFT+U method following the approach of Dudarev
\emph{et al}~\cite{Dudarev1998}. Several on-site Hubbard U parameters ($U =1.0, 1.5, 2.0$ eV) are tested for Hf($5d$) orbitals, which yield almost the same results as that from the GGA calculations (see Supporting Information). To study the topological edge states, we construct the maximally localized Wannier functions using the Wannier90 code~\cite{Marzari1997,Souza2001}, and then calculate the edge states using the iterative Green's function method as implemented in the Wannier$\_$tools package~\cite{Wu_tool}.

\section{RESULTS}\label{section:results}
\begin{figure}[t!]
\centerline{\includegraphics[width=0.5\textwidth]{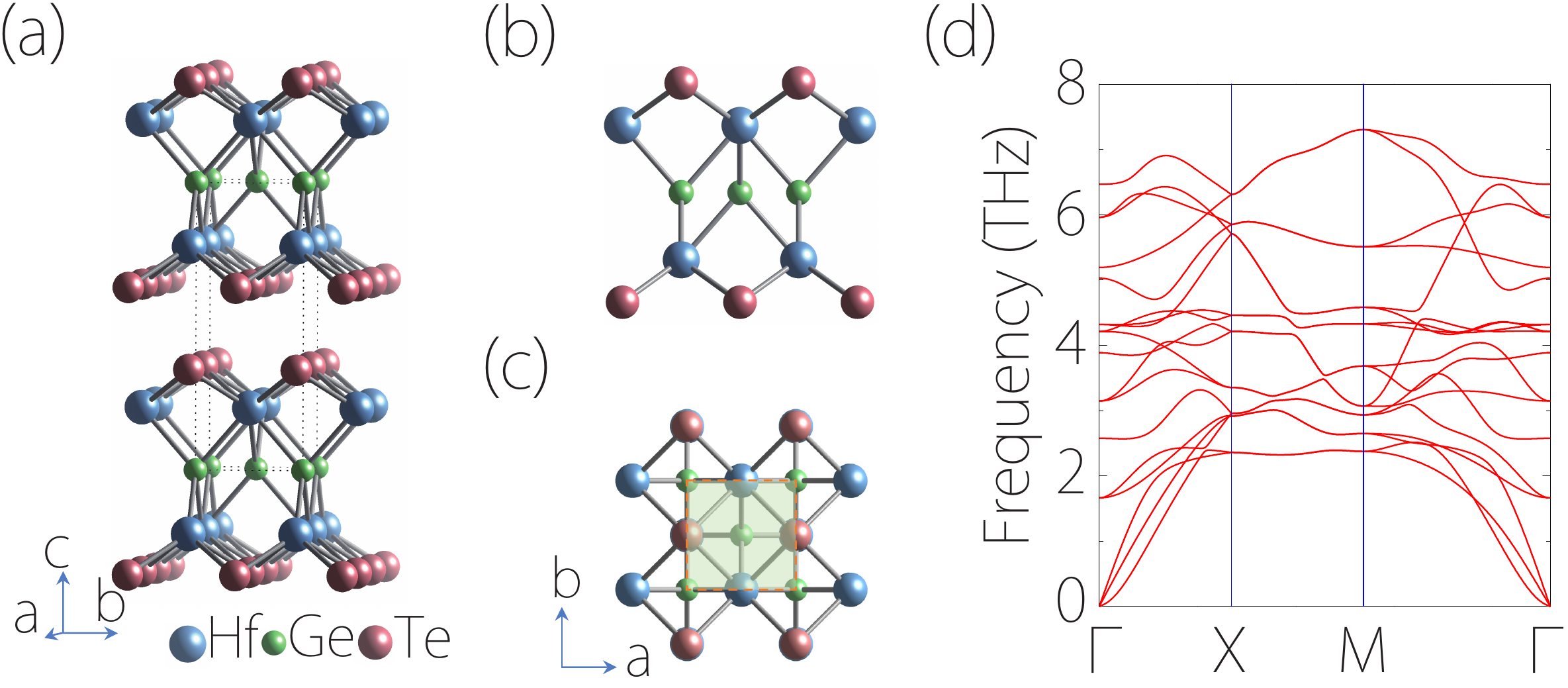}}
\caption{(a) Crystal structure of three-dimensional bulk HfGeTe. Here two HfGeTe layers are shown. The dotted lines indicate the unit cell for the bulk structure. (b) Side view and (c) top view of 2D monolayer HfGeTe (ML-HfGeTe). The shaded area in (c) indicates the unit cell for the monolayer. (d) Phonon spectrum of ML-HfGeTe, showing the dynamical stability of the structure. }\label{fig1}
\end{figure}

In their three-dimensional (3D) bulk form, the HfGeTe-family materials take the PbFCl-type structure with space group No. 129 ($P4/nmm$) (see Fig.~\ref{fig1}(a))~\cite{Onken1964}. In fact, this structure is shared by a large group of (more than 200) existing compounds with chemical formula $WHM$, with $W$ a transition metal or rare earth element, and $H$/$M$ two main group elements~\cite{Wang1995}. Some of these materials have been gaining interest in recent research. For example, Dirac lines have been investigated by several works in the 3D bulk ZrSi$X$ ($X=$ S, Se, Te)~\cite{Schoop2016,Hu2016,Neupane2016,Topp2016,Hosen2017,Singha2017} and HfSiS materials~\cite{Takane2016,Chen2017}, and some materials like ZrSiO have been predicted to be topological insulators in the monolayer form~\cite{Xu2015}.  In this work, we shall consider the family with $W=\mathrm{Hf}$, $M=$Te, and $H$ being an element from the carbon group (group IVA). These materials all share similar electronic properties, so we shall mainly focus on HfGeTe as a representative in the following discussion.

As shown in Fig.~\ref{fig1}(a), the 3D bulk HfGeTe crystal has a layered structure with each HfGeTe layer consisting of five atomic layers in the sequence of Te-Hf-Ge-Hf-Te, and each atomic layer is of a square lattice. The lattice parameters obtained from DFT calculations are given by $a = b = 3.885$ \AA\, and $c = 8.464$ \AA, which is in good agreement with experimental values ($a = b = 3.87$ \AA\, and $c = 8.50$ \AA)~\cite{Onken1964}.

In this work, our focus is on the monolayer HfGeTe (ML-HfGeTe). The corresponding lattice structure is shown in Fig.~\ref{fig1}(b) and \ref{fig1}(c). The monolayer possesses the same space group symmetry as the bulk. The fully relaxed structure has lattice parameters $a=b=3.727$ \AA, which are slightly decreased from the bulk values. To check the structural stability of ML-HfGeTe,  we have calculated the phonon spectrum. As shown in Fig.~\ref{fig1}(d), there is no imaginary frequency (soft mode) in the phonon spectrum throughout the Brillouin zone, indicating that the material is dynamically stable. One also notes that at low frequencies near $\Gamma$ point, apart from the linearly dispersing in-plane transverse acoustic modes, there is also the parabolic out-of-plane acoustic (ZA) branch, which is a characteristic feature of 2D materials~\cite{Liu2007,Zhu2014}.

We further estimate the capability of ML-HfGeTe to form a freestanding membrane. The elastic strength of a 2D material can be characterized by its in-plane stiffness constant $C=(1/A_0)(\partial^2 E_S/\partial \varepsilon^2)|_{\varepsilon=0}$, where $A_0$ is the equilibrium area, $E_S$ is the strain energy given by the energy difference between the strained and unstrained systems, and $\varepsilon$ is the applied uniaxial strain. From DFT calculations, we find that $C\approx 0.69$ eV/\AA$^2$ for ML-HfGeTe. Consider the deformation of a freestanding ML-HfGeTe flake under gravity, from the elastic theory by balancing gravity and 2D strain energy~\cite{Booth2008,Zhao2014}, we find the ratio between the out-of-plane deformation and the dimension of the flake to be as small as $10^{-3}$ to $10^{-4}$ even for large flakes with a size about $10^4$ $\mu$m$^2$. This result suggests that ML-HfGeTe could be strong enough to form a freestanding 2D structure even without support of a substrate.

\begin{figure}[t!]
\centerline{\includegraphics[width=0.46\textwidth]{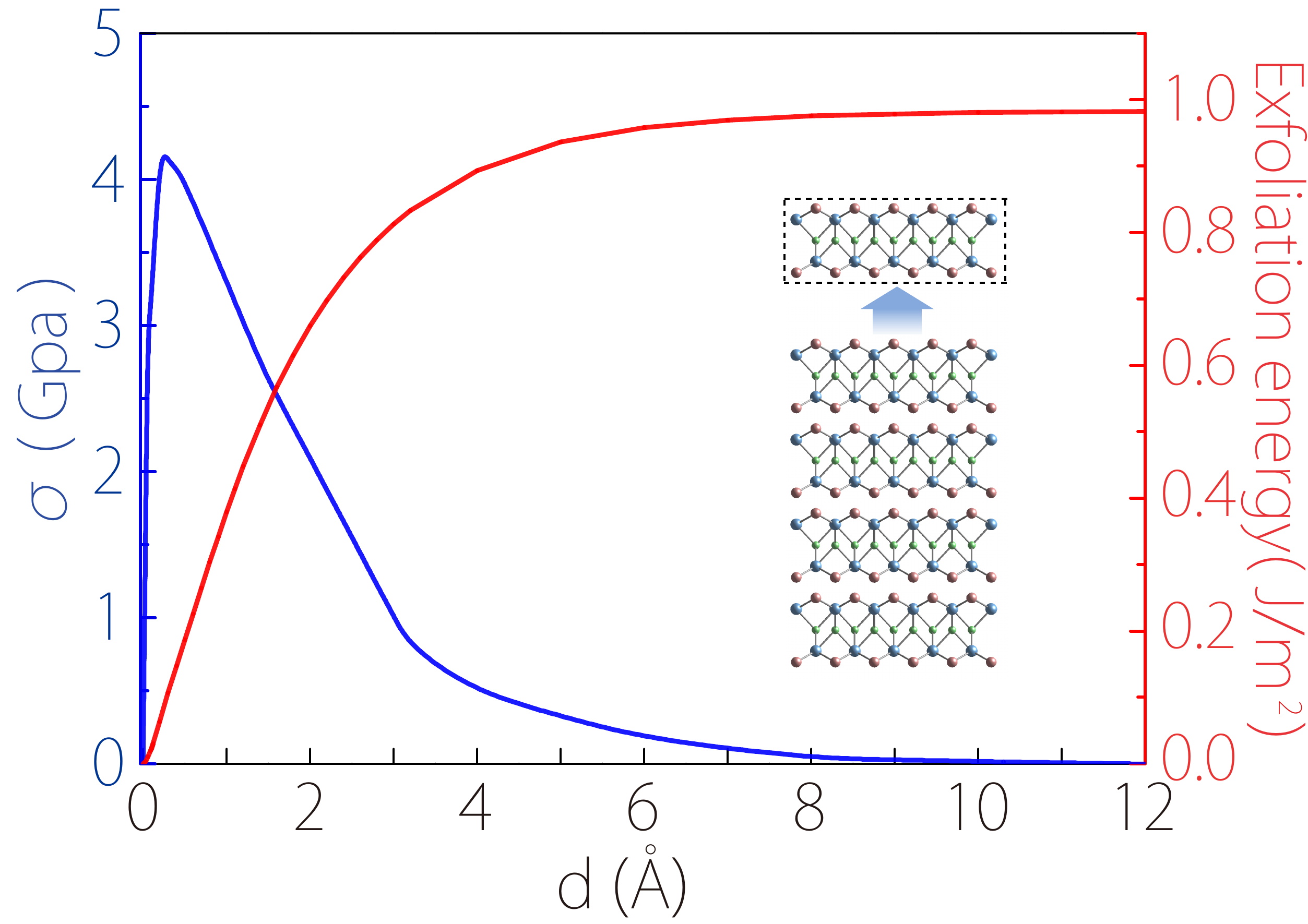}}
\caption{Calculated exfoliation energy (red line) for ML-HfGeTe as a function of its separation distance $d$ from the bulk (as illustrated in the inset). Here the bulk is modeled by four HfGeTe layers in the calculation, for which we have checked that the result converges. The blue curve shows the exfoliation strength $\sigma$ (i.e., the derivative of exfoliation energy with respect to $d$).}\label{fig2}
\end{figure}

Having confirmed the stability of ML-HfGeTe, we then check the possibility to obtain ML-HfGeTe from its bulk samples via mechanical exfoliation.
For bulk samples, the binding between the HfGeTe layer is relatively weak. In Fig.~\ref{fig2}, we plot the energy variation ($E_{ex}$) when a ML-HfGeTe is separated from the bulk by a distance $d$, simulating the exfoliation process. With increasing $d$, the energy quickly saturates to a value corresponding to the exfoliation energy about 0.98 J/m$^2$. This value is comparable to that of graphene ($\sim0.37$ J/m$^2$)~\cite{Zacharia2004} and MoS$_2$ ($\sim 0.41$ J/m$^2$) and is less than that of Ca$_2$N ($\sim 1.14$ J/m$^2$)~\cite{Zhao2014,Guan2015}, suggesting the feasibility to obtain ML-HfGeTe by mechanical exfoliation from the bulk. In Fig.~\ref{fig2}, we also show the exfoliation strength $\sigma$, which is obtained as the maximum derivative of $E_{ex}$ with respect to the separation distance $d$. The calculated value of $\sigma$ is about 4.2 GPa, also similar to values for typical 2D materials, such as graphene ($\sim 2.1$ GPa). In fact, mechanical exfoliation of ultrathin layers of the closely related materials ZrSiSe and ZrSiTe have already been demonstrated in experiment~\cite{Hu2016}.

\begin{figure*}[!htb]
\begin{center}
\centerline{\includegraphics[width=0.8\textwidth]{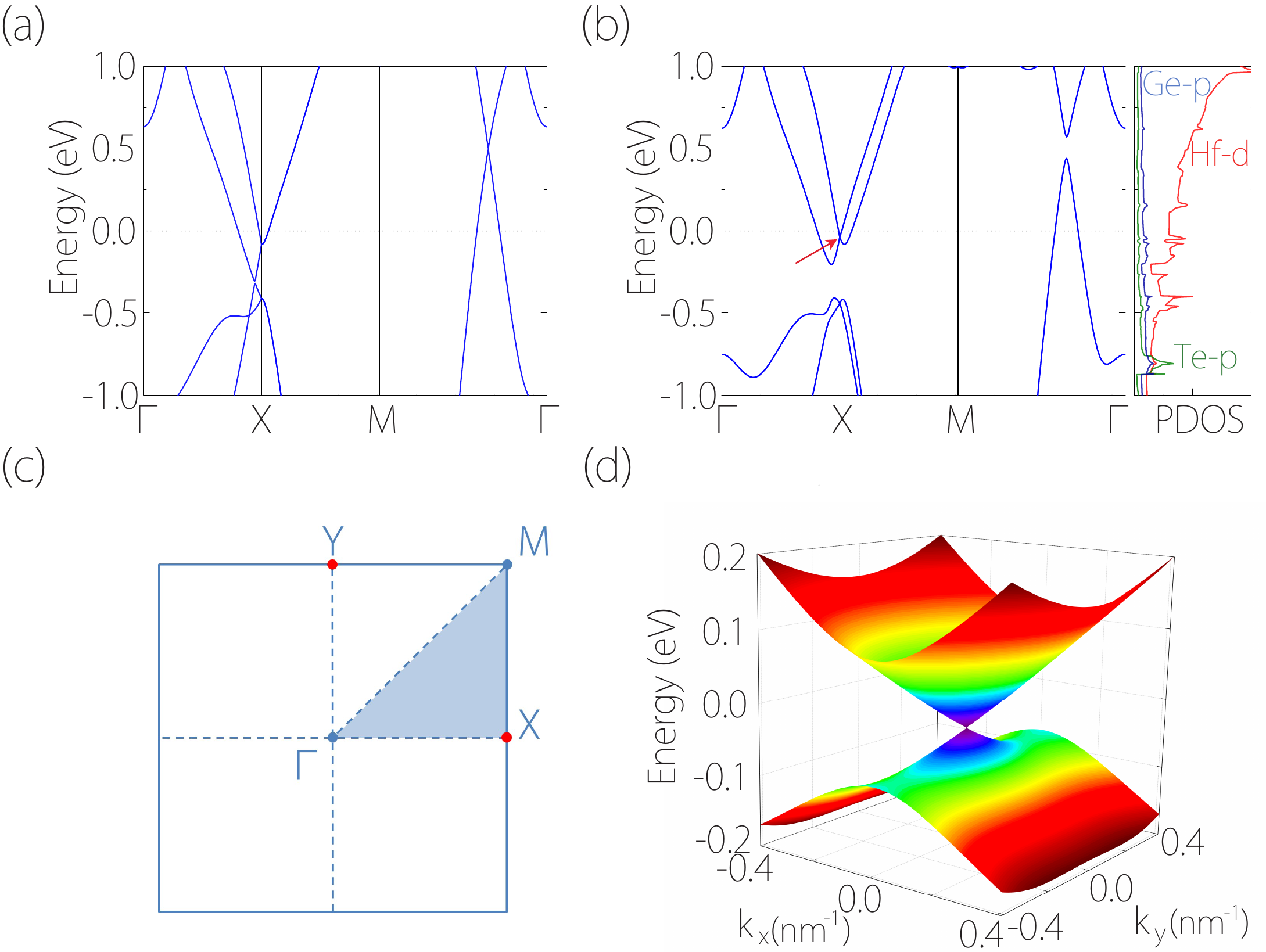}}
\caption{ (a) Electronic band structure of ML-HfGeTe without SOC. (b) Band structure of ML-HfGeTe with SOC included. The right panel shows the projected density of states (PDOS). The red arrow indicates the spin-orbit Dirac point (SDP) in the spectrum. (c) 2D Brillouin zone of ML-HfGeTe, with high symmetry points labeled. The shaded region indicates the irreducible Brillouin zone. The two low-energy SDPs are located at X and Y points, as indicated by the red dots. (d) Energy dispersions around the SDP at X, showing an anisotropic Dirac-cone spectrum.}\label{fig3}
\end{center}
\end{figure*}

Next, we come to the electronic properties of ML-HfGeTe. Figure~\ref{fig3}(a) and \ref{fig3}(b) show the electronic band structures of ML-HfGeTe without and with SOC, respectively. In the absence of SOC, the system exhibits a metallic state, with several band-crossings observed around X point and along the $\Gamma$-M path. By projecting the states onto atomic orbitals, we find that the low-energy states near the Fermi level are mainly from the Hf-$5d$ orbitals, which should have a large SOC effect. Indeed, after including SOC, the low-energy bands undergo noticeable changes. First, SOC opens gap at the original band-crossings around the X point and along the $\Gamma$-M path. However, a global bandgap is not opened, so the state remains a metal, with electron pocket near X point and hole pocket on $\Gamma$-M. Second, the original band-degeneracy along X-M path is lifted. Importantly, one observes Dirac-like dispersion with the Dirac point exactly located at X point and close to the Fermi level (at $-38$ meV). In addition, one notes that the structure preserves inversion symmetry, thus each band is at least two-fold degenerate. Hence the crossing point at X is four-fold degenerate, conforming with the definition of a Dirac point.

To confirm the linear dispersion around this point, in Fig.~\ref{fig3}(d), we plot the 2D energy surface around X point for the two crossing bands. One indeed observes Dirac-type linear dispersion along all directions from the point. The obtained Fermi velocities are anisotropic due to the absence of four-fold rotational symmetry at X, with $v_1=6.64\times 10^5$ m/s along $\Gamma$-X direction, and $v_2=3.86\times 10^5$ m/s along X-M direction. This Dirac point exhibits the unusual feature that it is robust under strong SOC (actually it only appears when SOC is included), and it is located at the time-reversal-invariant momentum (TRIM) point on the BZ boundary. These observations suggest that it is a SDP.

To fully demonstrate its identity as a 2D SDP, we need to clarify the mechanism that protects the Dirac point against SOC. In the following analysis, the electron spin and the SOC are included. We shall show that the presence of SDP at X point is solely dictated by the following symmetries of the system: glide mirror plane $\widetilde{\mathcal{M}}_z:(x,y,z)\rightarrow (x+1/2,y+1/2,-z)$, inversion $\mathcal{P}$, and time reversal symmetry $\mathcal{T}$. Here $\widetilde{\mathcal{M}}_z$ is a nonsymmorphic symmetry, which involves translation of half the lattice parameter along both $x$ and $y$ directions. Since every $k$-point in BZ is invariant under $\widetilde{\mathcal{M}}_z$ operation, each Bloch state can be chosen as eigenstate of $\widetilde{\mathcal{M}}_z$. To find the possible $\widetilde{\mathcal{M}}_z$ eigenvalues, we note that
\begin{equation}
(\widetilde{\mathcal{M}}_z)^2=T_{110}\overline{E}=-e^{-ik_x-ik_y},
\end{equation}
where $T_{110}$ denotes the translation by one unit cell along both $x$ and $y$ directions, and $\overline{E}$ is a $2\pi$ rotation on spin which leads to a $(-1)$ factor (here the wave-vectors $k_x$ and $k_y$ are measured in unit of $1/a$). Hence the eigenvalues of $\widetilde{\mathcal{M}}_z$ are given by
\begin{equation}\label{gz}
g_z=\pm i e^{-ik_x/2-ik_y/2}.
\end{equation}
It is important to note that the inversion does not commute with the glide mirror operation, instead, one finds
\begin{equation}\label{commun}
\widetilde{\mathcal{M}}_z\mathcal{P}=T_{110}\mathcal{P}\widetilde{\mathcal{M}}_z=e^{-ik_x-ik_y}\mathcal{P}\widetilde{\mathcal{M}}_z.
\end{equation}
As a result, for an eigenstate $|g_z\rangle$ of $\widetilde{\mathcal{M}}_z$ with eigenvalue $g_z$, the following relation holds:
\begin{equation}\label{eq3}
\widetilde{\mathcal{M}}_z(\mathcal{PT}|g_z\rangle)=-g_z(\mathcal{PT}|g_z\rangle).
\end{equation}
As we have mentioned, due to the presence of both $\mathcal{P}$ and $\mathcal{T}$ symmetries, with SOC, each band here is (at least) two-fold degenerate with the pair of states $|g_z\rangle$ and $\mathcal{PT}|g_z\rangle$ at each $k$-point. Now Eq.~(\ref{eq3}) shows that the two states $|g_z\rangle$ and $\mathcal{PT}|g_z\rangle$ have \emph{opposite} $\widetilde{\mathcal{M}}_z$ eigenvalues. At the special point X with $k_x=\pi$ and $k_y=0$, the relations in Eqs.~(\ref{gz}) and (\ref{commun}) are reduced to $g_z=\pm 1$ and $\{\widetilde{\mathcal{M}}_z,\mathcal{P}\}=0$. Further note that X is a TRIM point, hence any state $|g_z\rangle$ at X has another degenerate Kramers partner $\mathcal{T}|g_z\rangle$ with the same eigenvalue $g_z$ (because $g_z=\pm 1$ is real at X). This ensures four-fold degeneracy at X point, with the following linearly independent states $\{ |g_z\rangle, \mathcal{T}|g_z\rangle, \mathcal{PT}|g_z\rangle, \mathcal{P}|g_z\rangle$\}.  Note that for a generic $k$-point deviating from X, it will \emph{not} be invariant under $\mathcal{T}$ operation, hence the four-fold degeneracy will generally split into two doubly-degenerate bands away from X point. Thus, an isolated four-fold band-crossing point must appear at X. From the band structure in Fig.~\ref{fig3}(b), one indeed observes that the states at X are all of this type, such as the crossing-point below the local gap. In the symmetry analysis, spin is explicitly considered, so the resulting degeneracy is robust against SOC.

\begin{figure*}[!htb]
\centerline{\includegraphics[width=0.9\textwidth]{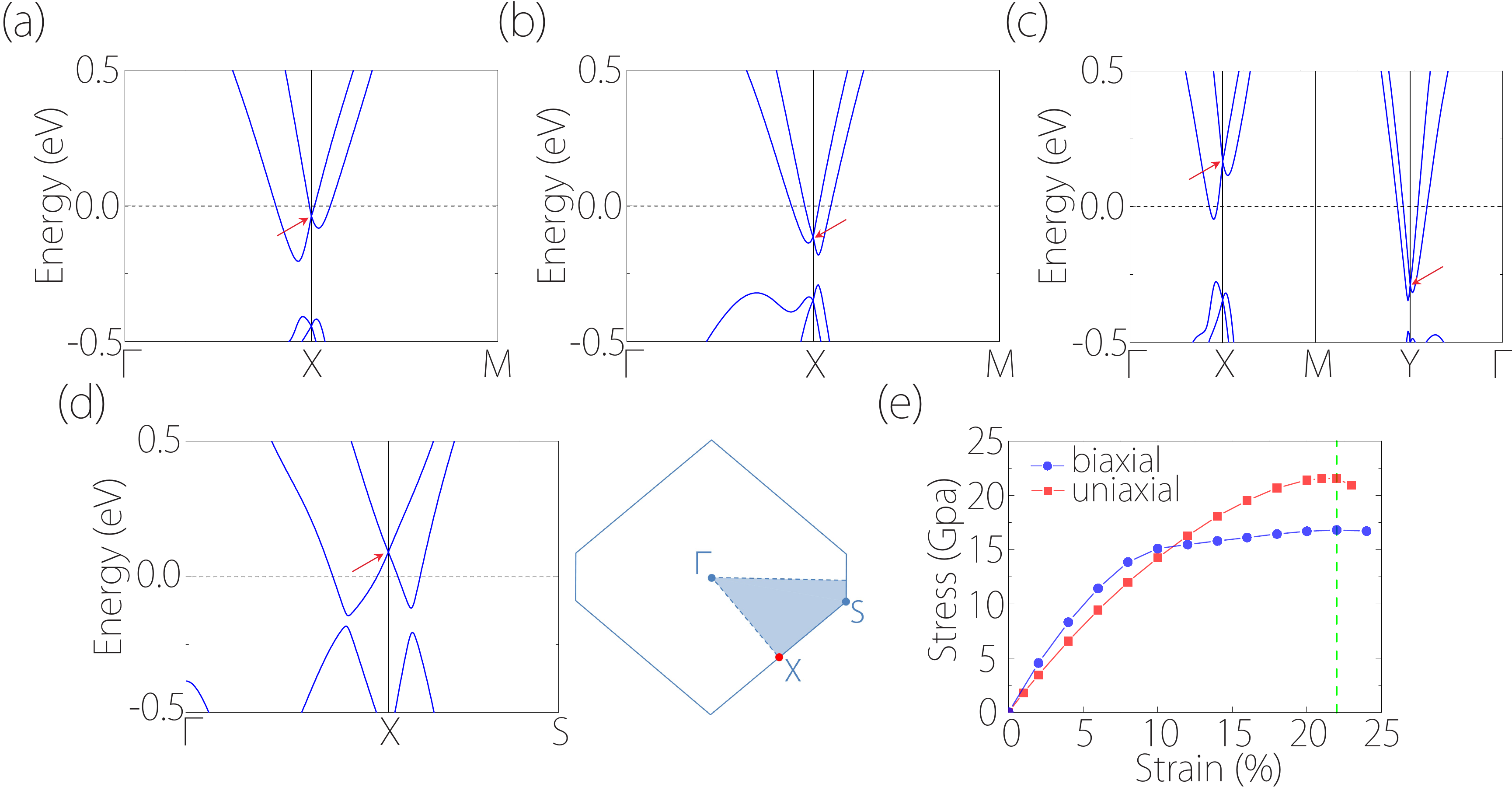}}
\caption{ Band structure of ML-HfGeTe under lattice distortion: (a) without strain; (b) with $+$5\% biaxial strain; (c) with $+$5\% uniaxial strain; and (d) varying the angle between $a$ and $b$ to $80^{\circ}$. SDPs are preserved in all these cases, as indicated by the red arrows. For uniaxial strain in (c), the X and Y points are no longer related by symmetry, so they behave independently. The right panel in (d) shows the Brillouin zone for the distorted lattice structure. (e) Strain-stress relations for ML-HfGeTe with different types of strains. The vertical dashed line marks the approximate critical tensile strain (about 22\%). }\label{fig4}
\end{figure*}

 In the group-theory language, the SDPs here derive from the existence of four-fold irreducible representations of the nonsymmorphic space group at X point. For $k$-points on the BZ boundary, these representations are obtained from projective representations of the associated crystal point group (using Herring's method), which can be related to the regular representations of a larger group known as the central extension group~\cite{Bradley1972}. This typically gives rise to higher dimensional irreducible representations, leading to band degeneracy points on the BZ boundary.

The SDP and the Dirac fermions around it are characterized by the effective $k\cdot p$ Hamiltonian. This model can be obtained from the symmetry constraints that the Hamiltonian $\mathcal{H}$ is invariant under the symmetry operations at X~\cite{Steinberg2014}. The symmetry operations in the little group at X include $\mathcal{T}$, $\mathcal{P}$, $\widetilde{\mathcal{M}}_z$, and additionally $\mathcal{M}_y: (x,y,z)\rightarrow (x,-y+1/2,z)$, which is an (off-centered) mirror plane perpendicular to $y$.  The matrix representations of these operations can be found in the standard reference~\cite{Bradley1972}, with $\mathcal{T}=-i\sigma_y\otimes \tau_0 K$, $\mathcal{P}=\sigma_0\otimes\tau_x$, $\widetilde{\mathcal{M}}_z=\sigma_0\otimes\tau_z$, $\mathcal{M}_y=-i\sigma_y\otimes \tau_x$. Here $K$ is the complex conjugation, $\sigma_i$ and $\tau_i$ ($i=x,y,z$) are the Pauli matrices representing the degree of freedom with the four degenerate states at X, $\sigma_0$ and $\tau_0$ are the $2\times 2$ identity matrix.
Constrained by these symmetries, the effective Hamiltonian expanded around the 2D SDP can be expressed as
\begin{equation}
\mathcal{H}(\bm k)=v_1 k_x(\cos\theta\sigma_z\otimes \tau_z+\sin\theta \sigma_x\otimes \tau_z)+v_2 k_y\sigma_y\otimes \tau_z,
\end{equation}
where the wave-vector $\bm k$ and the energy are measured from the SDP, and the model parameters $v_1$, $v_2$, and $\theta$ are real and depend on the microscopic details. The obtained energy dispersion is given by $E=\pm \sqrt{v_1^2k_x^2+v_2^2 k_y^2}$,  where each eigenvalue is doubly degenerate due to the combined $\mathcal{PT}$ symmetry, consistent with the anisotropic Dirac-cone spectrum that one observes in Fig.~\ref{fig3}(d).

Up to this point, we have established the presence of SDP in ML-HfGeTe. It should be pointed out that Y point at $(0,\pi)$ is connected to X point by a four-fold rotational symmetry (see Fig.~\ref{fig3}(c)), hence there is also a SDP at Y with its Dirac-cone rotated by $\pi/2$ compared to X. Thus, ML-HfGeTe is a 2D spin-orbit Dirac material with a pair of SDPs close to its Fermi level.

Since the SDP in ML-HfGeTe is protected by the $\mathcal{T}$, $\mathcal{P}$, and $\widetilde{\mathcal{M}}_z$ symmetries, the Dirac point cannot be destroyed as long as these symmetries are preserved. We find that these symmetries are quite robust: they survive under a variety of strains, such as in-plane biaxial, uniaxial, and shear strains. In Fig.~\ref{fig4}, we plot the calculated band structures under several different strains. One indeed observes that the SDP is maintained for all these cases, only the energy of the Dirac point and the dispersion are changed by strain. The strain-stress curves in Fig.~\ref{fig4}(e) show that ML-HfGeTe also has excellent mechanical properties. It exhibits a linear elastic region up to $8\%$ strain, and the critical strain is beyond 20\%. These suggest that strain can be employed as an effective way to tune the properties of spin-orbit Dirac fermions in ML-HfGeTe.

\begin{figure}[!htb]
\centerline{\includegraphics[width=0.5\textwidth]{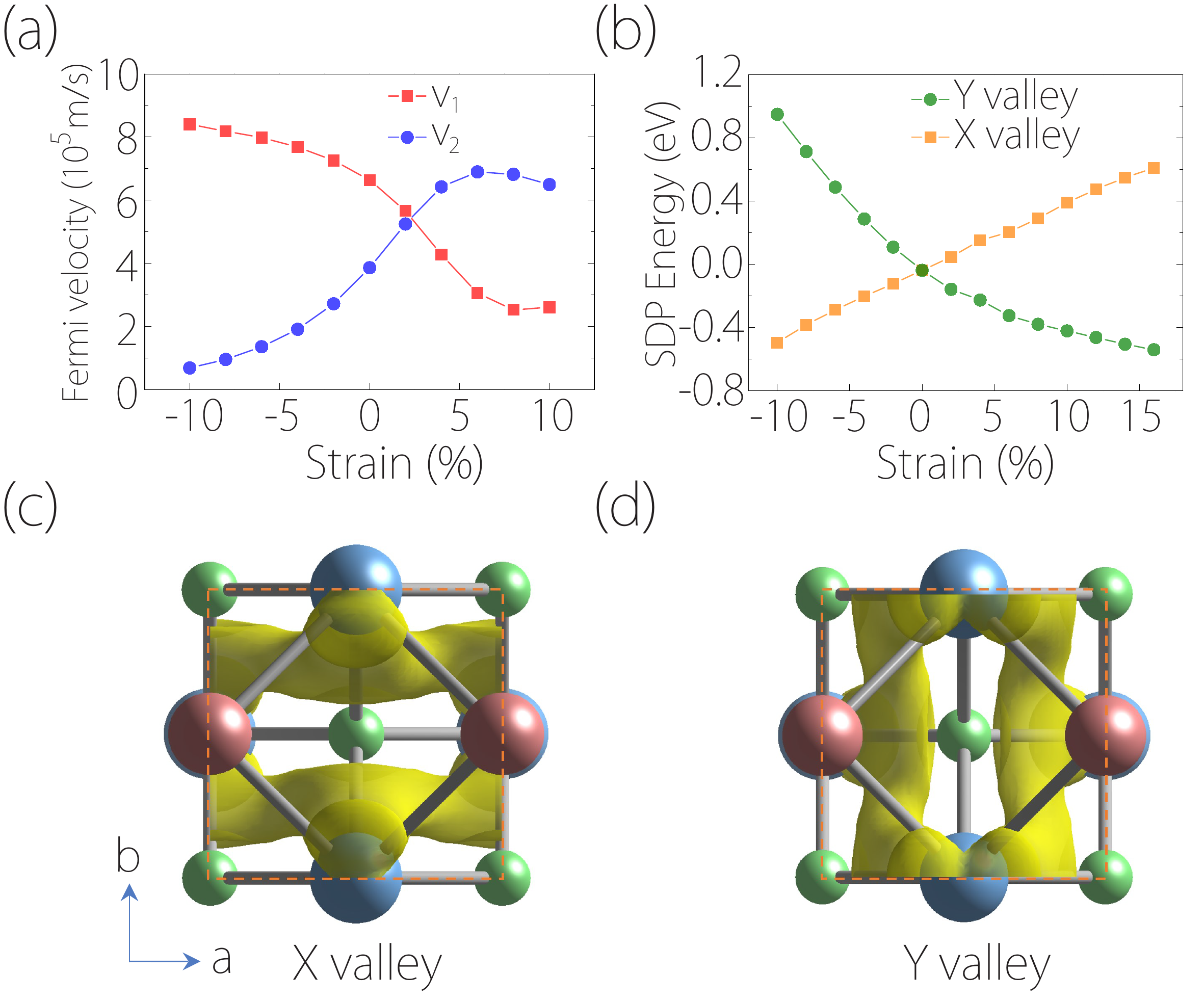}}
\caption{(a) Values of the two Fermi velocities $v_1$ and $v_2$ at the SDP versus the applied biaxial strain. (b) Energies of the SDPs versus the applied uniaxial strain. Here the two curves are for the two SDPs at X valley and Y valley, respectively. (c,d) Charge density distribution plotted for states at (c) X valley and (d) Y valley.}\label{fig5}
\end{figure}

For example, the Fermi velocities can be tuned by strain. As plotted in Fig.~\ref{fig5}(a), under biaxial strain, the two Fermi velocities $v_1$ and $v_2$ can be changed on the order of $5\times 10^5$ m/s in the range between $-5\%$ to $+5\%$ strains. The case with uniaxial strain is even more interesting, because in this case, the four-fold rotational symmetry that connects X and Y points are broken. Consequently, the two valleys at X and Y become independent. In Fig.~\ref{fig4}(c), one observes that the two SDPs are shifted in opposite directions along energy axis. This behavior can be understood from the different bonding features at X and Y valleys. In Fig.~\ref{fig5}(c) and \ref{fig5}(d), we plot the charge distribution for states at X and Y valleys. One observes that the states at X valley shows a bonding character along $x$, whereas the states at Y valley shows a bonding character along $y$ (as dictated by the four-fold rotational symmetry). Therefore, when stretched along $x$, the X valley will be pushed up in energy, while the Y valley will be shifted down due to contraction along $y$. This is consistent with the result in Fig.~\ref{fig4}(c). The variation of SDP energies  versus uniaxial strain along $x$-direction is shown in Fig.~\ref{fig5}(b). The energy separation between the two SDPs can be up to 0.6 eV at 5\% strain. We find that above $\sim6\%$ uniaxial strain, the valley at X is above the Fermi level and becomes unoccupied, resulting in a large valley polarization with Dirac fermions all in the Y valley.

It is known that a $\mathbb{Z}_2$-classification applies for 2D insulators with preserved time-reversal symmetry, where a nontrivial $\mathbb{Z}_2$ invariant indicates a 2D topological insulator phase~\cite{Kane2005}. Interestingly, we find that ML-HfGeTe also possesses a nontrivial $\mathbb{Z}_2$ invariant. Here, although ML-HfGeTe does not have a global bandgap, one notes that its bandgap is closed \emph{indirectly}, i.e., there is no direct bandgap closing at any $k$-point [see Fig.~\ref{fig3}(b)]. Thus, the band structure can be adiabatically connected to a fully-gapped insulating phase without any band crossing in the process (e.g., by raising the conduction bands at X point and by lowering the valence bands along the $\Gamma$-M path), such that it is also characterized with a $\mathbb{Z}_2$ invariant. Here, the $\mathbb{Z}_2$ invariant is defined for the bands below the \emph{local} gap. Such an idea was theoretically proposed before in Ref.~\onlinecite{Pan2014} based on a model study, in which the proposed 2D metallic phase with a nontrivial $\mathbb{Z}_2$ invariant was termed as a 2D $\mathbb{Z}_2$ topological metal.

\begin{figure}[!htb]
\centerline{\includegraphics[width=0.5\textwidth]{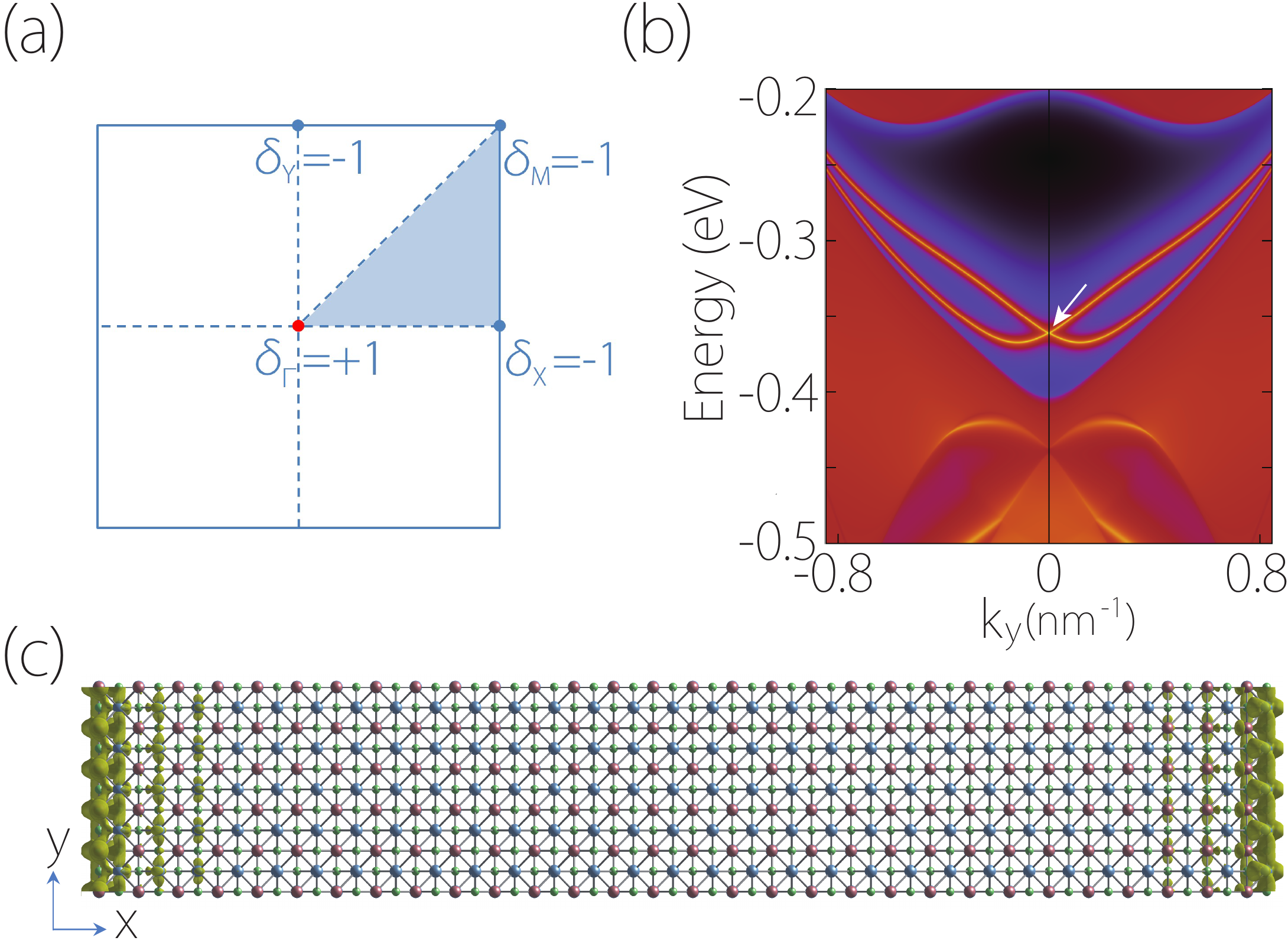}}
\caption{(a) Product of band parity eigenvalues at the four TRIM points for ML-HfGeTe. (b) Surface energy spectrum for a semi-infinite ML-HfGeTe, with a single boundary running along $y$-direction. Topological edge states can be observed. (c) Charge density distribution for edge state, obtained from the DFT calculation of a ML-HfGeTe ribbon with a width of 30 unit cells along $x$-direction. Here the plot is for the edge states at $k_y=0$ (corresponding to that indicated by the arrow in (b)). Note that the ribbon is taken to be infinite along $y$, here we only show a section of it.}\label{fig6}
\end{figure}

We have rigorously evaluated the $\mathbb{Z}_2$ invariant using the formula derived by Fu and Kane for centrosymmetric systems~\cite{Fu2007}. In this approach, one analyzes the parity eigenvalues at the four TRIM points ($\Gamma$, X, Y, and M). At each TRIM point, we calculate the quantity $\delta_i=\prod_{m=1}^N \xi_{2m}(i)$, where $i\in\{\Gamma$, X, Y, M\}, $\xi_{2m}(i)=\pm 1$ is the parity eigenvalue of the $2m$th band at $i$, which shares the same eigenvalue $\xi_{2m}=\xi_{2m-1}$ with its Kramers degenerate partner, and $m$ runs through the $2N$ bands below the local gap. Then the $\mathbb{Z}_2$ invariant $\nu=0,1$ is obtained from the product of the four $\delta_i$'s through~\cite{Fu2007}
\begin{equation}
(-1)^\nu=\prod_i \delta_i.
\end{equation}
We have calculated the $\delta_i$'s using our DFT result, and their values are listed in Table~\ref{tab:tab1} [also see Fig.~\ref{fig6}(a)]. We indeed find that the $\mathbb{Z}_2$ invariant $\nu=1$ is nontrivial for ML-HfGeTe. Thus, ML-HfGeTe also serves as the first realistic example that realizes the 2D $\mathbb{Z}_2$ topological metal phase proposed in Ref.~\onlinecite{Pan2014}.

\begin{table}[h]
\caption{$\mathbb{Z}_2$ invariant $\nu$ of ML-HfGeTe evaluated from parity analysis.}
\label{tab:tab1}
\begin{tabular}{p{2.4cm}<{\centering}|p{1cm} p{1cm} p{1cm} p{1cm}}
  \hline\hline
  $\mathbb{Z}_2$ invariant $\nu$ & $\delta_\Gamma$ & $\delta_\text{X}$ & $\delta_\text{Y}$ & $\delta_\text{M}$ \\ \hline
    1 & +1 & -1 & -1 & -1 \\
  \hline\hline
\end{tabular}
\end{table}

The nontrivial $\mathbb{Z}_2$ invariant dictates the presence of topological edge states at the sample boundaries~\cite{Kane2005a,Hasan2010,Qi2011}. We consider a semi-infinite system of ML-HfGeTe with a boundary along $y$-direction. The edge spectrum is calculated and plotted in Fig.~\ref{fig6}(b). One indeed observes a pair of topological edge states around the edge-projected X point (at $k_y=0$), similar to those for 2D topological insulators. These states are spin-helical, i.e., the states (at the same energy) propagating along opposite directions are with opposite spin polarizations. To further confirm the localization of these states around the edge, we directly calculate the charge density distribution of the edge state for a ML-HfGeTe nano-ribbon. The result in Fig.~\ref{fig6}(c) shows that the edge state is indeed confined at the sample edge. The existence of these topological edge states manifests the nontrivial topology of the bulk band structure of ML-HfGeTe.

\section{Discussion}

We emphasize that the existence of SDPs here is solely dictated by the nonsymmorphic space group symmetry. Hence they also appear in other member materials of the HfGeTe-family which share the same crystal symmetry, such as ML-HfSnTe and ML-HfSiTe etc. (see Supporting Information) and also materials with Te substituted by other chalcogen elements. However, symmetry cannot constrain the energy of the SDP. Fortunately, we find that for all the members of this family, the SDPs appear close to the Fermi level, hence qualifying them as 2D SDP materials. This provides a number of candidate materials that can be used to explore 2D SDP fermions.

We mentioned that these proposed materials represent the first realistic material platform that realizes the 2D SDPs proposed by Young and Kane~\cite{Young2015a}. In this family of materials, there are two symmetry equivalent SDPs (at X and Y) connected by the four-fold rotation, which corresponds to Case I discussed in Ref.~\onlinecite{Young2015a}. When the rotational symmetry is broken by the lattice strain [as in Fig.~\ref{fig4}(c)], the two SDPs become inequivalent, which then corresponds to Case II in Ref.~\onlinecite{Young2015a}.

Several experimental and computational works have found Dirac lines in 3D bulk materials ZrSi$X$ ($X=$S, Se, Te) and HfSiS~\cite{Schoop2016,Hu2016,Neupane2016,Topp2016,Hosen2017,Singha2017,Takane2016,Chen2017}. The Dirac lines there are also derived from nonsymmorphic symmetries, similar to our case. But there are important distinctions. (i) Those works are on 3D bulk materials, whereas our work focuses on 2D materials. Note that this dimensionality difference is crucial for the stabilization of Dirac points, as we mentioned in the Introduction. (ii) The materials studied in those works have relatively weak SOC, such that the bands along certain paths (X-M and R-A in the 3D BZ) become nearly degenerate, forming the Dirac lines. In comparison, the ML-HfGeTe studied here has stronger SOC, such that the bands well split except at the X and Y points, leading to well-defined 2D SDPs.

As we have mentioned, the 3D bulk material of HfGeTe (and of HfSiTe) already exist. This would greatly facilitate the realization of 2D monolayers, e.g., by using the mechanical exfoliation method. Experimentally, it has been demonstrated that for two closely related materials ZrSiSe and ZrSiTe, ultrathin layers with thickness less than 10 nm can be readily obtained by mechanical exfoliation~\cite{Hu2016}. Therefore, we expect that ML-HfGeTe (and other materials in the family) could also be fabricated in the near future. Once realized, the Dirac dispersion can be directly probed via angle-resolved photoemission spectroscopy (ARPES). The Dirac fermion character typically would tend to enhance the carrier mobility. The topological edge states can be detected via ARPES or scanning tunneling spectroscopy at the sample edge.

In this work, we focus on the Dirac points which have linear energy dispersions. There could exist other kinds of protected band degeneracy points with  quadratic or higher-order dispersions. For example, 2D quadratic band-touching points have been found in bilayer graphene~\cite{McCann2006} and in blue phosphorene oxide~\cite{Zhu2016}. It will be interesting to explore these exotic band degeneracy points and their novel physics in future studies.

Finally, we point out that the HfGeTe-family is still not ideal in terms of manifesting the Dirac physics, because besides the SDP, there are also extraneous non-Dirac bands passing the Fermi level. They will lead to additional contributions in electronic properties. Thus, in future research, it will be desirable to explore new materials with clean Dirac band structures, and/or explore methods to engineer the band structure of existing materials to get rid of the extraneous bands. Nevertheless, our current work serves an important first step towards this goal.

\section{CONCLUSION}\label{section:conclusions}
In conclusion, based on first-principles calculation and theoretical analysis, we have proposed the first realistic material that realizes 2D SDPs close to its Fermi level. This material, ML-HfGeTe, is shown to be stable in monolayer form, and may be easily exfoliated from the corresponding bulk material. There exists a pair of SDPs in ML-HfGeTe. From symmetry analysis, we identify the nature and protection of the SDPs, demonstrating that they are intrinsically robust against SOC. We construct an effective $k\cdot p$ model for characterizing the low-energy fermions around the SDP. It is shown that these SDPs are quite robust against lattice deformations, and various strains can be used as powerful means to tune the SDPs. Furthermore, we find that ML-HfGeTe also represents the first example of the previously proposed 2D $\mathbb{Z}_2$ topological metal. It possesses a nontrivial $\mathbb{Z}_2$ invariant in the bulk, and a pair of topological edge states on the boundary. The above features are shared by a number of 2D materials in the HfGeTe-family. Our findings thus provide a promising platform to explore the intriguing physics of 2D spin-orbit Dirac fermions and the associated nanoscale applications.

\begin{acknowledgments}
The authors thank D.L. Deng for helpful discussions. This work is supported by the MOST Project of China (Grants No.2016YFA0300603 and No.2014CB920903), the National Natural Science Foundation of China (Grants No.11574029), the Singapore Ministry of Education Academic Research Fund Tier 2 (MOE2015-T2-2-144) and Tier 1 (SUTD-T1-2015004).We acknowledge computational support from Texas Advanced Computing Center.

\end{acknowledgments}

%

\end{document}